\documentclass[11pt,a4paper]{article}
\pdfoutput=1

\usepackage{jheppub}
\usepackage{booktabs}
\usepackage{siunitx}

\newcommand{\trace}[1]{\left<{#1}\right>}

\title{Electroweak Skyrmions in the HEFT}

\author{Juan Carlos Criado, Valentin V. Khoze and Michael Spannowsky}

\affiliation{Institute for Particle Physics Phenomenology, Department of Physics, Durham University, Durham, DH1 3LE, UK}

\emailAdd{juan.c.criado@durham.ac.uk}
\emailAdd{valya.khoze@durham.ac.uk}
\emailAdd{michael.spannowsky@durham.ac.uk}

\abstract{
We study the existence of skyrmions in the presence of all the electroweak degrees of freedom, including a dynamical Higgs boson, with the electroweak symmetry being non-linearly realized in the scalar sector. For this, we use the formulation of the Higgs Effective Field Theory (HEFT). In contrast with the linear realization, a well-defined winding number exists in HEFT for all scalar field configurations. We classify the effective operators that can potentially stabilize the skyrmions and numerically find the region in parameter spaces that support them. We do so by minimizing the static energy functional using neural networks. This method allows us to obtain the minimal-energy path connecting the vacuum to the skyrmion configuration and calculate its mass and radius. Since skyrmions are not expected to be produced at colliders, we explore the experimental and theoretical bounds on the operators that generate them. Finally, we briefly consider the possibility of skyrmions being dark matter candidates.
}

\begin{document}

\maketitle

\section{Introduction}

Skyrmions are extended field configurations that behave as new particle degrees of freedom. Initially, they were proposed as a description of baryons within an Effective Field Theory (EFT) description of strong interactions containing only the pion fields~\cite{Skyrme:1961vq, Witten:1979kh, Adkins:1983ya}. Since the pions can be viewed as pseudo-Goldstone bosons arising from the breaking of an $SU(2)$ symmetry, the original setting can be directly applied to the electroweak sector, in the limit in which the Higgs field is infinitely massive, and the gauge bosons are decoupled, so that only the $SU(2)$ would-be Goldstone bosons are present~\cite{Ellis:2012cs}. Electroweak skyrmions have been shown to survive under certain conditions in more realistic settings, in which these limits are partially removed, which can destroy the topological protection they enjoyed in the first place~\cite{Ambjorn:1984bb}.

The purpose of ref.~\cite{Criado:2020zwu} and of this work is to consider skyrmions in the full electroweak theory, including the effects of both the gauge fields and a dynamical Higgs boson. As in the original Skyrme setting, the existence of skyrmions in the Standard Model (SM) Lagrangian is forbidden by Derrick's theorem~\cite{Derrick:1964ww}, but they can be stabilized by including higher-order effective operators. Since the discovery of the Higgs~\cite{ATLAS:2012yve, CMS:2012qbp}, two effective descriptions of the electroweak sector have emerged: the Standard Model EFT (SMEFT), in which the scalars furnish a linear representation of the electroweak symmetry group; and the Higgs Effective Field Theory (HEFT), in which the realization of this symmetry is non-linear. The SMEFT version is studied in ref.~\cite{Criado:2020zwu}. In this paper, we focus on the HEFT framework, which we find to be better suited for the description of skyrmions because of the non-trivial topology of its scalar sector.

In section~\ref{sec:theory}, we briefly introduce the relevant sector of the HEFT, discuss the differences with the SMEFT and with the approximations that have been previously taken, and introduce the topological numbers that characterize the topology of its field configurations. In section~\ref{sec:results}, we study the existence of skyrmions numerically in the presence of the different combinations of HEFT operators. In section~\ref{sec:pheno}, we consider the phenomenological consequences of skyrmions and the operators that generate them. This allows us to obtain constraints on the parameter space, in which we include positivity bounds. We summarize our conclusions in section~\ref{sec:conclusions}.

\section{Theory}
\label{sec:theory}

The relevant degrees of freedom for skyrmions in the electroweak sector are the $SU(2)$ gauge bosons $W^a_\mu$, the would-be Goldstone bosons $G^a$, and the Higgs boson $h$. We neglect the effects of the $U(1)_Y$ gauge sector. The Higgs is invariant under $SU(2)$ gauge transformations, while the Goldstones are collected in a non-linear representation
\begin{equation}
 U = \exp \frac{i \sigma^a G^a}{\sqrt{2} v} \in SU(2),
\end{equation}
with no relation to the Higgs singlet field $h$.\footnote{This is to be contrasted with the more restrictive linear realization where $h(x)$ and $U(x)$ are assembled into the Higgs doublet $ \phi = \frac{1}{\sqrt{2}}\left(v +h\right) U
 \begin{pmatrix}
  0 \\ 1
 \end{pmatrix}$.}

We write the effective Lagrangian as
\begin{equation}
 \mathcal{L} = \sum_i F_i(h/v) \mathcal{Q}_i,
 \qquad \qquad
 F_i(\eta) = \sum_{n=0}^\infty c_{i,n} \eta^n,
\end{equation}
where $\Lambda$ is the HEFT cut-off scale; the $\mathcal{Q}_i$ are monomials in $W_{\mu\nu}$, $U$, $h$ and their covariant derivatives, with $h$ appearing only through its derivatives. That is, schematically
\begin{equation}
 \mathcal{Q}_i \sim h^{\mathrm{H}_i} (W_{\mu\nu})^{\mathrm{W}_i} U^{\mathrm{U}_i} D^{\mathrm{D}_i},
\end{equation}
where $\mathrm{H}_i$, $\mathrm{W}_i$, $\mathrm{D}_i$ are, respectively, the number of Higgs fields, field-strength tensors, and covariant derivatives contained in $\mathcal{Q}_i$. In this setting, $\mathrm{D}_i$ corresponds to the general chiral dimension~\cite{Buchalla:2018yce}.

We adopt a power counting based on the chiral dimension, in which each $c_{i,n}$ coefficient is of order $\Lambda^{2 - \mathrm{D}_i}$, multiplied by the necessary power of $v$ for the coefficient to have the correct energy dimensions. Thus, terms with higher chiral dimensions are suppressed by higher powers of $v/\Lambda$.

We keep terms with chiral dimension 4, and impose custodial symmetry, which is needed for configurations in the spherical ansatz to give spherically symmetric contributions to the energy, as described in section~\ref{sec:spherical-ansatz}. A list of all relevant operators $\mathcal{Q}_i$ is given in table~\ref{tab:operators}, partially following the notation of ref.~\cite{Buchalla:2013rka}, where angle brackets $\trace{\cdot}$ denote a trace and $L_\mu = i U D_\mu U^\dagger$.

\begin{table}
 \centering
 \begin{tabular}{ccc}
  \toprule
  Name
  & Operator
  & Radial energy density $\rho_i$ in spherical ansatz
  \\
  \midrule
  $\mathcal{Q}_1$ & 1 & $-\frac{r^2}{e^2 v^4}$
  \\
  $\mathcal{Q}_h$ & $\partial_\mu h \partial^\mu h$ & $\frac{r^2}{2} (\eta')^2$
  \\
  $\mathcal{Q}_U$ & $\trace{D_\mu U^\dagger D^\mu U}$
  & $\frac{2}{v^2} \left(f_1^2 + f_2^2 + \frac{r^2}{2} b^2\right)$
  \\
  $\mathcal{Q}_{X h 2}$ & $\trace{W_{\mu\nu} W^{\mu\nu}}$
  & \small $-8 e^2 \Big[
     (f_1' - 2 b f_2)^2 + (f_2' + (2 f_1 - 1) b)^2
     + \frac{2}{r^2} (f_1^2 + f_2^2 - f_1)^2 
     \Big]$
  \\
  $\mathcal{Q}_{X h 5}$ & $\epsilon^{\mu\nu\rho\sigma} \trace{W_{\mu\nu} W_{\rho\sigma}}$
  & $0$
  \\
  $\mathcal{Q}_{X U 8}$ & $i\trace{W_{\mu\nu} [L^\mu, L^\nu]}$
  & \small $\frac{16 e^2}{2 r^2} \Big[
     (f_1^2 + f_2^2)(f_1^2 + f_2^2 - f_1 + 2 r^2 b^2)
     - b r^2 (f_2 f_1' - f_1 f_2' + b f_1)
   \Big]$
  \\
  $\mathcal{Q}_{X U 11}$ & $i\epsilon^{\mu\nu\rho\sigma} \trace{W_{\mu\nu} [L_\rho, L_\sigma]}$
  & $0$
  \\
  $\mathcal{Q}_{D1}$ & $\trace{L_\mu L^\mu}^2$
  & $-\frac{4 e^2}{r^2} \left[2 (f_1^2 + f_2^2) + r^2 b^2\right]^2$
  \\
  $\mathcal{Q}_{D2}$ & $\trace{L_\mu L_\nu} \trace{L^\mu L^\nu}$
  & $-\frac{4 e^2}{r^2} \left[2 (f_1^2 + f_2^2)^2 + r^4 b^4\right]$
  \\
  $\mathcal{Q}_{D7}$ & $\trace{L_\mu L^\mu} \partial_\nu h \partial^\nu h$
  & $-e^2 v^2 (\eta')^2 \left[2 (f_1^2 + f_2^2) + r^2 b^2\right]$
  \\
  $\mathcal{Q}_{D8}$ & $\trace{L_\mu L_\nu} \partial^\mu h \partial^\nu h$
  & $- e^2 v^2 (\eta')^2 r^2 b^2$
  \\
  $\mathcal{Q}_{D11}$ & $(\partial_\mu h \partial^\mu h)^2$
  & $-\frac{e^2 v^4}{4} (\eta')^4 r^2$
  \\
  \bottomrule
 \end{tabular}
 \caption{Custodial-invariant operators $\mathcal{Q}_i$ containing the Higgs only through derivatives, of order up to $\Lambda^0$, together with their contribution to the radial energy density $\rho_i$ in the spherical ansatz, defined in eq.~\eqref{eq:radial-density}. \label{tab:operators}}
\end{table}

The relevant sector of the SM Lagrangian is given by the chiral dimension 2 operators, with
\begin{gather}
 F_1(h/v) = V(h) = \lambda v^4 \left((h/v)^2 + (h/v)^3 + \frac{1}{4}(h/v)^4\right) = \lambda \left(v^2 h^2 + v h^3 + \frac{h^4}{4}\right),
 \\
 F_h(h/v) = \frac{1}{2},
 \qquad
 F_{Xh2}(h/v) = -\frac{1}{2 g^2},
 \qquad
 F_U(h/v) = \frac{v^2}{4} \left(1 + \frac{h}{v}\right)^2.
\end{gather}
Deviations from the SM are encoded in modifications of any of the $F_i(h)$. Derrick's theorem forbids the existence of solitons in the SM. A necessary condition for them to exist is that higher-derivative terms are present. The original term proposed by Skyrme~\cite{Skyrme:1961vq} to stabilize skyrmions can be written in the HEFT Lagrangian as 
\begin{equation}
 \mathcal{L}_{\mathrm{Sk}} =
 -\, \frac{1}{16 e^2} (\mathcal{Q}_{D1} - \mathcal{Q}_{D2}),
 \label{eq:skyrme-term}
\end{equation}
that is, setting $F_{D1}(h/v) = -F_{D2}(h/v) = -\,1 / (16 e^2)$. In the chiral dimension power-counting, the size of the coefficient is given by $e \sim \Lambda / (4 v)$. The theory $\mathcal{L}_{\text{SM}} + \mathcal{L}_{\text{Sk}}$ is thus a candidate for the stabilization of skyrmions. Two limits of it have been previously studied in the literature:
\begin{enumerate}
\item[\textbf{A.}] Frozen Higgs. This corresponds to $m_h \to \infty$, which implies that the Higgs is set to its vev $h = 0$ everywhere.
\item[\textbf{B.}] No gauge fields. This is obtained when the $SU(2)$ vanishes, $g \to 0$. In this limit, the coefficient of the $\trace{W_{\mu\nu} W^{\mu\nu}}$ term becomes large, and the gauge fields are forced to approach a pure gauge configuration in order to minimize the energy. One can gauge them away. The only degrees of freedom left are the Goldstone bosons and the Higgs.
\end{enumerate}
Taking both limits leads to a theory with only the Goldstone bosons as dynamical degrees of freedom, which has been studied in, e.g. ref.~\cite{Adkins:1983ya}. Limit \textbf{A} has been considered in ref.~\cite{Ambjorn:1984bb}, while limit \textbf{B} has been considered in ref.~\cite{Kitano:2016ooc}. In any of these limits, and in the full theory, the Skyrme term can be generalized by allowing other linear combinations of the $\mathcal{Q}_{D1}$ and $\mathcal{Q}_{D2}$ operators. This has been done in the case where both limits are taken, in ref.~\cite{Ellis:2012cs}, and in limit \textbf{B}, in ref.~\cite{Kitano:2017zqw}.

\medskip

In ref.~\cite{Criado:2020zwu} skyrmions were studied in the full theory without assuming any of the two limits above. This was done within the SMEFT framework, in which the electroweak symmetry is realized linearly. The purpose of the present paper is to continue this program in the non-linear realization. Ultimately, the existence of skyrmions turns out to be much harder to prove in the SMEFT than in the HEFT, as discussed below. Ref.~\cite{Hamada:2021oqm}, which appeared during the preparation of this work, has a similar scope.

\medskip

In limit \textbf{B}, the theory contains stable field configurations separated from the vacuum by an infinite energy barrier. This fact can be understood from a topological point of view. 
To have \emph{finite} energy, the scalar fields must satisfy the following boundary conditions:
\begin{equation}
 \lim_{|\mathbf{x}| \to \infty} h(x) = 0, \qquad
 \lim_{|\mathbf{x}| \to \infty} U(x) = 1_{2 \times 2},
\end{equation}
which means that all directions towards infinity can be identified with a single point, effectively compactifying space into $S^3$. Thus, the fields can be viewed as a $S^3 \to \mathbb{R} \times S^3$ mapping. We can then define a topological charge, the winding number for the $U : S^3 \to S^3$ part of the mapping:
\begin{equation}
 n_U = \frac{1}{24 \pi^2} \epsilon_{ijk} \int d^3x \trace{L_i L_j L_k}.
\end{equation}
This is a homotopy invariant of $U$, and therefore it can never change with smooth time evolution. However, this number is only well defined when the target space of the scalar sector has the topology $\mathbb{R} \times S^3$. This is true generically both in the full HEFT and in limit \textbf{B}, but it ceases to be in the particular case of the SMEFT, in which the Lagrangian becomes independent of $U(x)$ when $h(x) = -v$ (see footnote 1 above). One can then identify all points with this value of $h$, turning the scalar manifold into $\mathbb{R}^4 \cong \mathbb{C}^2$. The scalar degrees of freedom are thus collected into a $SU(2)$ doublet $\phi$. In general, the topology of the static configurations of $\phi$ cannot be characterized in terms of the number $n_U$ since $U$ is only defined through 
$ \phi = \frac{1}{\sqrt{2}}\left(v +h\right) U\cdot 
 (0\,1)^T$ when $\phi \neq 0$ everywhere.\footnote{Even if $\phi = 0$ only at an isolated point $p$, $U$ becomes a mapping $S^3 - \{p\} \cong \mathbb{R}^3 \to S^3$, and all such mappings are homotopically equivalent.} One can recover a well-defined $n_U$ in the SMEFT by taking the frozen Higgs limit \textbf{A}, which disallows $h(x) = -v$ and forces the scalars to be in the submanifold $S^3$. As already noted earlier, we will not follow this route in this paper and will instead use the HEFT formulation of the theory where $h$ and $U$ are independent and without imposing limits \textbf{A} and \textbf{B}.

\medskip

The inclusion of gauge fields destroys the topological protection of $n_U \neq 0$ configurations from decaying into the vacuum. However, the $\trace{W_{ij} W^{ij}}$ term in the energy induces a finite-energy barrier between configurations in which $W_\mu$ is a pure gauge, 
$W_i = \mathcal{U} \partial_i \mathcal{U}^\dagger$, $W_0=0$,
possibly making them metastable. In order to describe this, we use the Chern-Simons number
\begin{equation}
 n_{\text{CS}} = \frac{1}{16 \pi^2} \epsilon_{ijk} \int d^3x \trace{
  W_i W_{jk} + \frac{2i}{3} W_i W_j W_k
 }.
\end{equation}
For a pure-gauge $W_i = \mathcal{U} \partial_i \mathcal{U}^\dagger$, $n_{\text{CS}}$ is the integer winding number of the gauge transformation $\mathcal{U} (\mathbf{x}): S^3 \to S^3$.

A skyrmion is a field configuration for which $n_U$ and $n_{\text{CS}}$ differ by (approximately\footnote{Due to metastability.}) one unit. We thus define the skyrmion number as
\begin{equation}
 n_{\text{Sk}} = n_U - n_{\text{CS}}.
\end{equation}
While $n_U$ and $n_{\text{CS}}$ are not gauge invariant, $n_{\text{Sk}}$ is, because $n_U$ and $n_{\text{CS}}$ change by the same integer under a large gauge transformation. An anti-skyrmion is similarly a configuration where $n_{\text{Sk}} \simeq -1$, and multi-skyrmions have $|n_{\text{Sk}}| > 1$. A CP transformation changes the sign of the skyrmion number.

\section{Skyrmion configurations and energy landscape}
\label{sec:results}

\subsection{The energy functional in the spherical ansatz}
\label{sec:spherical-ansatz}

We parametrize the space of static configurations of the fields $W^a_\mu$, $U$ and $h$ in the $W_0=0$ gauge by means of 4 real functions of one variable: $f_1$, $f_2$, $b$ and $\eta$. We do so by further imposing the unitary gauge $U (\mathbf{x})= 1_{2 \times 2}$ and the spherical ansatz:
\begin{equation}
 W_i(\mathbf{x}) = v e \tau_a \left(
  \epsilon_{ija} n_j \frac{f_1(r)}{r}
  + (\delta_{ia} - n_i n_a) \frac{f_2(r)}{r}
  + n_i n_a b(r)
 \right),
 \qquad
 h(\mathbf{x}) = \frac{v}{\sqrt{2}} \eta(r),
\end{equation}
where $\tau_a$ are the Pauli matrices, $n_i = x_i / |\mathbf{x}|$, $r = v e |\mathbf{x}|$, and $e$ is a parameter we will adjust as a function of Wilson coefficients. The energy density in this ansatz is spherically symmetric when all interactions are invariant under custodial symmetry.\footnote{Indeed, if one takes any $M \in SU(2)$ and $R$ its representation as a spatial rotation, one has 
$M W_i(\mathbf{x}) M^\dagger = R_{ij} \, W_j(R^{-1} \mathbf{x})$,
so invariance under spatial rotations and under custodial symmetry become equivalent.
}
We can then write the energy as
\begin{equation}
 E = \frac{4 \pi v}{e} \int_0^\infty dr \sum_i F_i(\eta) \rho_i,
 \label{eq:radial-density}
\end{equation}
where the contributions $\rho_i$ to the radial energy density of each $\mathcal{Q}_i$ operator are given in table~\ref{tab:operators}. Requiring that the energy is finite and that the fields are regular at the origin gives rise to the following boundary conditions:
\begin{gather}
 f_1(0) = f_1'(0) = f_2(0) = f_2'(0) - b(0) = \eta'(0) = 0, \\
 f_1(\infty) = f_2(\infty) = b(\infty) = \eta(\infty) = 0.
\end{gather}

Since we have fixed the unitary gauge, the skyrmion number is just $n_{\text{Sk}} = - n_{\text{CS}}$. For convenience, we define
\begin{equation}
 n_W = \frac{i}{24 \pi^2} \epsilon_{ijk} \int d^3x \trace{W_i W_j W_k}
 = \frac{2}{\pi} \int_0^\infty dr \; b (f_1^2 + f_2^2),
\end{equation}
which agrees with $n_{\text{CS}}$ at integer values. Thus, skyrmion and anti-skyrmions will be found at $n_W \simeq -1$ and $n_W \simeq 1$, respectively. CP symmetry, which takes one into the other, is given here by $f_1 \to f_1$, $f_2 \to - f_2$, $b \to -b$, $\eta \to \eta$. All the operators we consider are invariant under this transformation. This is because the two operators that violate CP vanish for static field configurations. Thus, the static-configuration energy functional is invariant under $n_W \to - n_W$.

\subsection{The Skyrme term}

\begin{figure}
 \centering
 \includegraphics[width=0.49\textwidth]{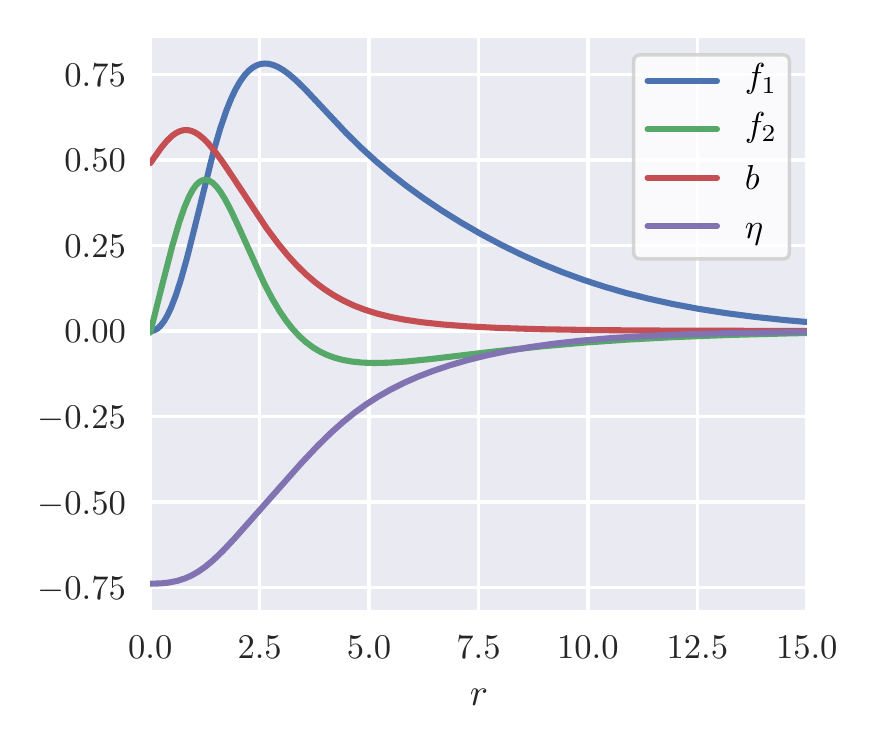}
 \includegraphics[width=0.49\textwidth]{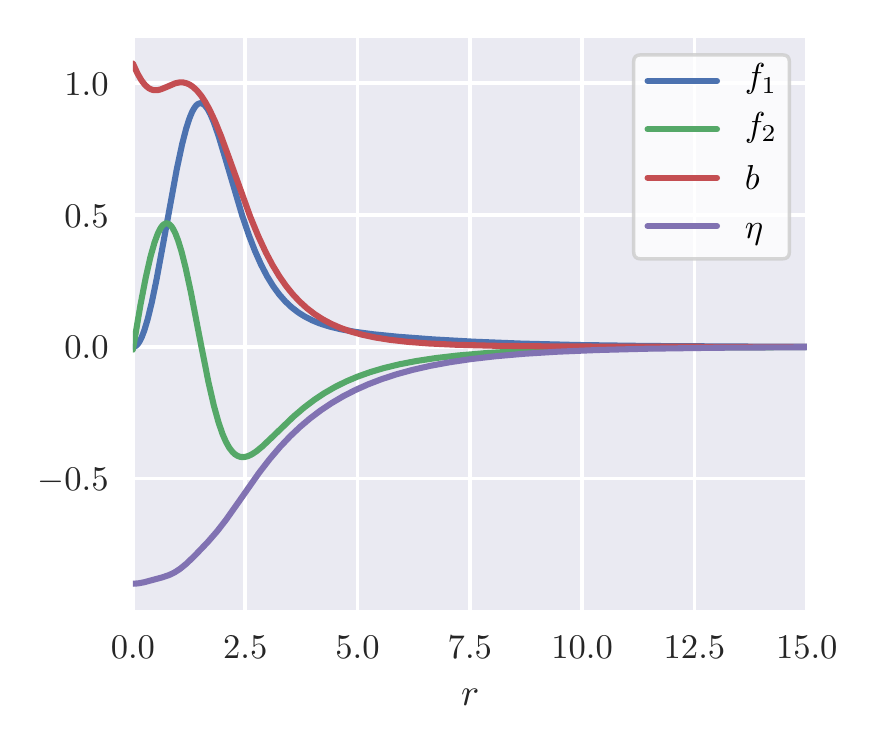}
 \caption{Minimal energy configurations for $e = 1.2$, $c_{D1,0} = -c_{D2,0} = 1/(16 e^2)$, and $n_W = 0.4$ (left) or $n_W = 0.8$ (right).}
 \label{fig:configs}
\end{figure}

\begin{figure}
 \centering
 \includegraphics[width=\textwidth]{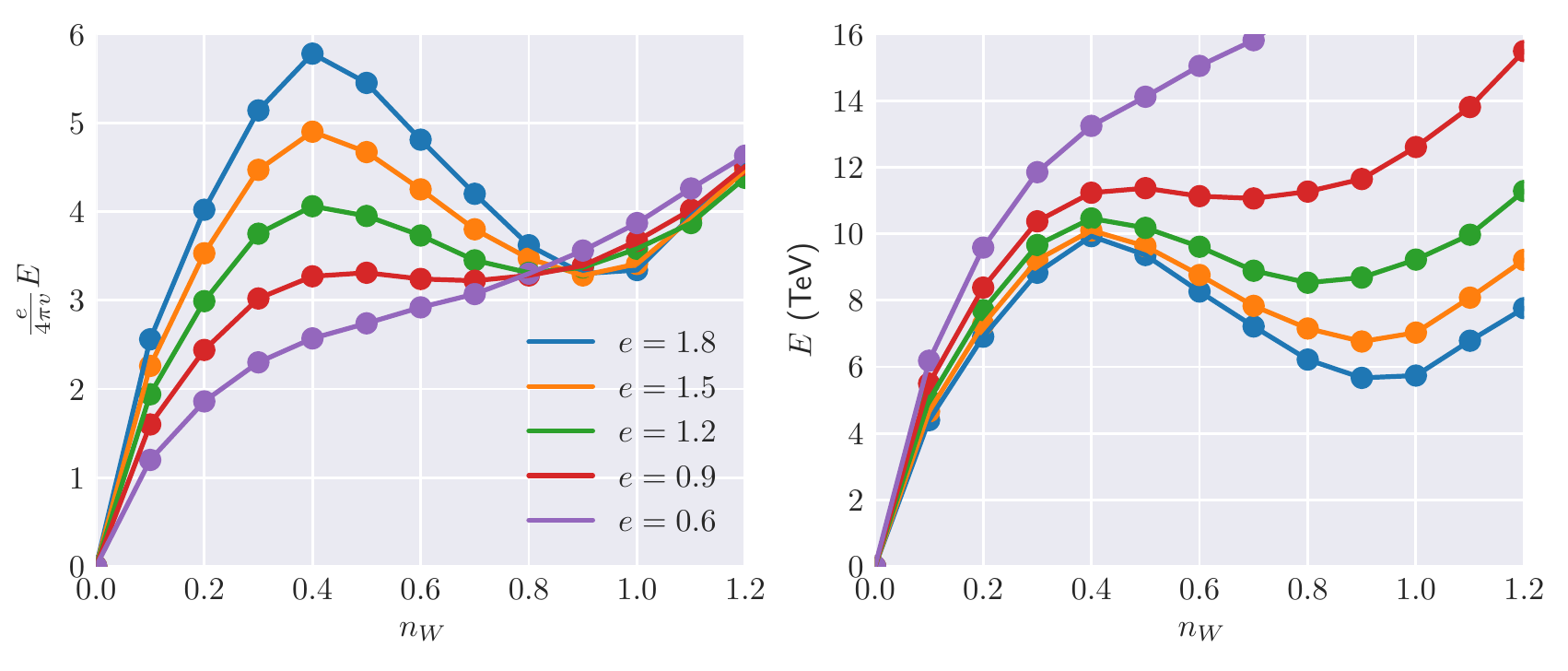}
 \caption{Minimal energy as a function of $n_W$, for $c_{D1,0} = -c_{D2, 0} = 1/(16 e^2)$ and different values of $e$. The finite-energy disappears around $e = 0.9$.}
 \label{fig:E-vs-nW}
\end{figure}

We focus first on the case in which
\begin{equation}
 -c_{D1,0} = c_{D2,0} \equiv \frac{1}{16 e^2},
 \label{eq:skyrme-coeffs}
\end{equation}
with the rest of non-SM coefficients in the HEFT Lagrangian being set to zero. The last equality is to be understood as fixing the free parameter $e$ of the ansatz. This corresponds to the original Skyrme term, given in eq.~\eqref{eq:skyrme-term}. The total energy functional is given by
\begin{equation}
 E = \frac{4 \pi v}{e} \int_0^\infty dr \left[
  \rho_{\text{SM}} + (f_1^2 + f_2^2) \left(b^2 + \frac{f_1^2 + f_2^2}{2 r^2}\right)
 \right],
\end{equation}
where $\rho_{\text{SM}}$ is the contribution from the SM. We shall now describe the field configurations and energy landscape that arise in this setting. We study them using the method described in appendix~\ref{sec:numerical-method}. We display two example configurations for $e = 1.8$ and different values of $n_W$ in figure~\ref{fig:configs}. In figure~\ref{fig:E-vs-nW}, we show the minimal energy as a function of $n_W$, for different values of $e$. For $e > e_{\text{crit}} \simeq 0.9$, we find a finite-energy barrier separating the skyrmion, with $n_W \simeq 1$, and the vacuum at $n_W = 0$. This barrier disappears below $e_{\text{crit}}$. Thus, the skyrmion solution exists only when $e > e_{\text{crit}}$ and is a metastable configuration. 

\begin{figure}
 \centering
 \includegraphics[width=0.6\textwidth]{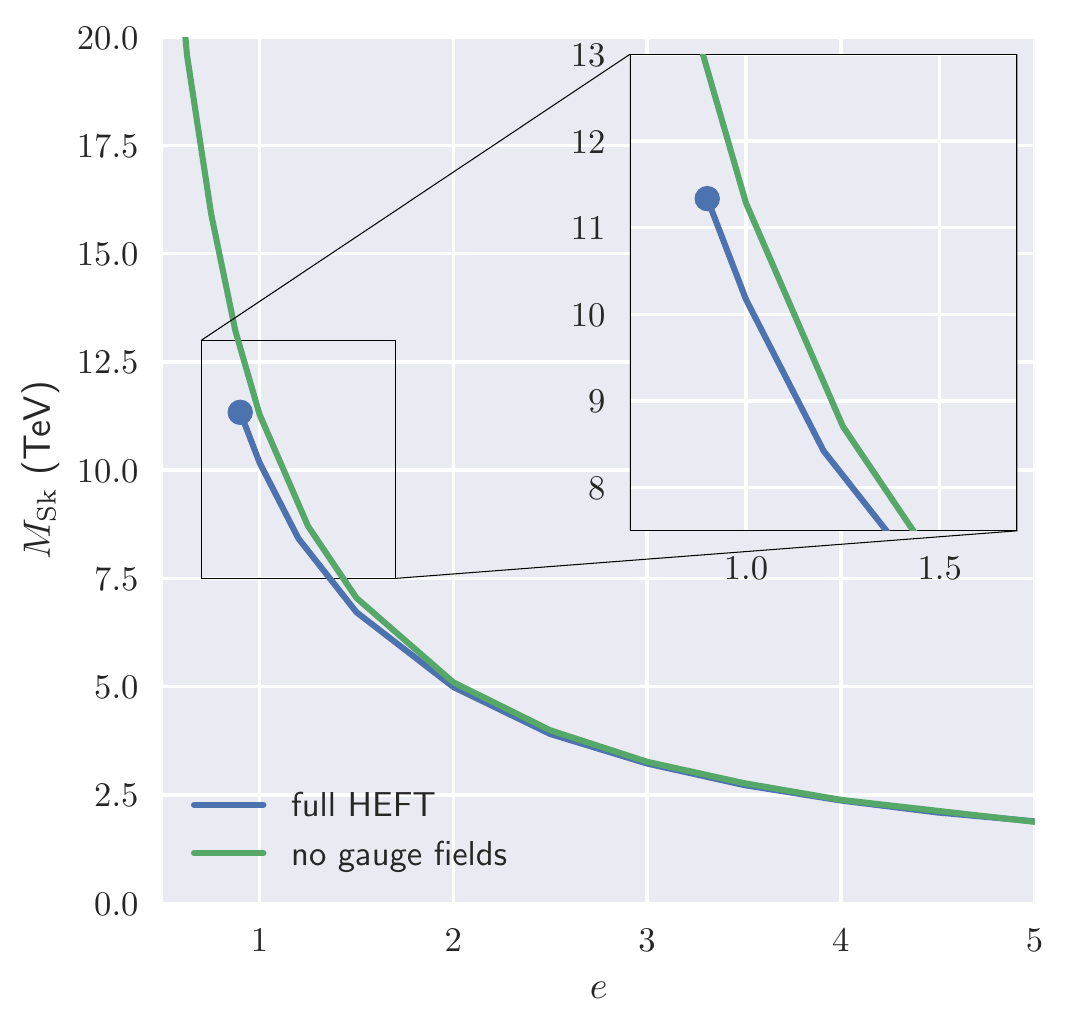}
 \caption{Skyrmion mass $M_{\text{Sk}}$ as a function of $e$ for the full theory and for limit \textbf{B}.}
 \label{fig:M-vs-e}
\end{figure}

The energy $E$ of the local minimum is the skyrmion mass. We find that the normalized energy $e M_{\text{Sk}} / (4 \pi v)$ is approximately constant, with a value of $3.3$ at $e = e_{\text{crit}}$, and a limiting value of $3$ as $e \to \infty$, so the skyrmion mass is given by
\begin{equation}
 M_{\text{Sk}}|_{e \simeq e_{\text{crit}}} \simeq \frac{41 v}{e},
 \qquad \qquad
 M_{\text{Sk}}|_{e \to \infty} \simeq \frac{38 v}{e}.
\end{equation}
The maximum of $M_{\text{Sk}}$ is reached at $e = e_{\text{crit}}$:
\begin{equation}
 M_{\text{Sk}} \leq M_{\text{Sk}}|_{e = e_{\text{crit}}} \simeq \SI{11}{TeV}.  
\end{equation}
In figure~\ref{fig:M-vs-e}, we show this behaviour and compare it to the case in which no gauge fields are present, labelled limit \textbf{B} in section~\ref{sec:theory}. The curves are similar for large $e$. This is to be expected since a large value of $e$ makes the $\trace{W_{\mu\nu} W^{\mu\nu}}$ term dominant, with similar effects as taking $g \to 0$, which is limit \textbf{B}. However, some differences arise at small $e$. Just above $e_{\text{crit}}$, the mass of the skyrmion in the full theory is slightly lower than in limit \textbf{B}. This is because the $n_W = 1$ is no longer topologically fixed in the full theory, and so $n_W$ can move to another value with lower energy. For $e < e_{\text{crit}}$, skyrmions become unstable in the full theory, but nothing changes in limit \textbf{B}, as they are still topologically protected.

For the height of the barrier, the energy of the local maximum near $n_W = 1/2$, we find that
\begin{equation}
 E_{\text{barrier}}|_{e = e_{\text{crit}}} \simeq \SI{11}{TeV},
 \qquad \text{and }
 E_{\text{barrier}}|_{e \to \infty} \simeq \SI{10}{TeV}.
\end{equation}

We also define the radius of the skyrmion $R_{\text{Sk}}$ by averaging over the $n_W$ density as
\begin{equation}
 R^2_{\text{Sk}}
 = \frac{i}{24 \pi^2} \epsilon_{ijk} \int d^3x \; |\mathbf{x}|^2 \, \trace{W_i W_j W_k}
 = \frac{2}{\pi (v e)^2} \int dr \; r^2 \, b \, (f_1^2 + f_2^2).
\end{equation}
We find that
\begin{equation}
 R_{\text{Sk}}|_{e = e_{\text{crit}}} \simeq \frac{1.4}{v e},
 \qquad \qquad
 R_{\text{Sk}}|_{e \to \infty} \simeq \frac{1.9}{v e},
\end{equation}

\medskip

\subsection{Skyrmion stabilisation from other operators in HEFT}

We consider here the possibility that skyrmions are stabilized by some operator from table~\ref{tab:operators} other than $\mathcal{Q}_{D1} - \mathcal{Q}_{D2}$. Some of these operators can be discarded for this purpose from general considerations: $\mathcal{Q}_1$, $\mathcal{Q}_h$ and $\mathcal{Q}_U$, by Derrick's theorem; and all operators containing a field-strength tensor can also be neglected since they vanish when the gauge fields are set to a pure gauge configuration. There are five remaining operators that can contribute: the $\mathcal{Q}_{Di}$ in table~\ref{tab:operators}.

We consider now turning on one $c_{Di,n}$ coefficient at a time while fixing the others to zero. We find that none of them are capable of stabilizing skyrmions except for $c_{D1,0}$ and $c_{D2,0}$. Indeed, for all the others, their radial energy density $\rho_i$ is multiplied by some monomial in $\eta$ or $\eta'$. One can then take $\eta = 0$ everywhere, which implies $F_i(\eta) \rho_i = 0$, and then skyrmions become unstable by Derrick's theorem. We have checked this numerically in several examples.

\begin{figure}
 \centering
 \includegraphics[width=0.6\textwidth]{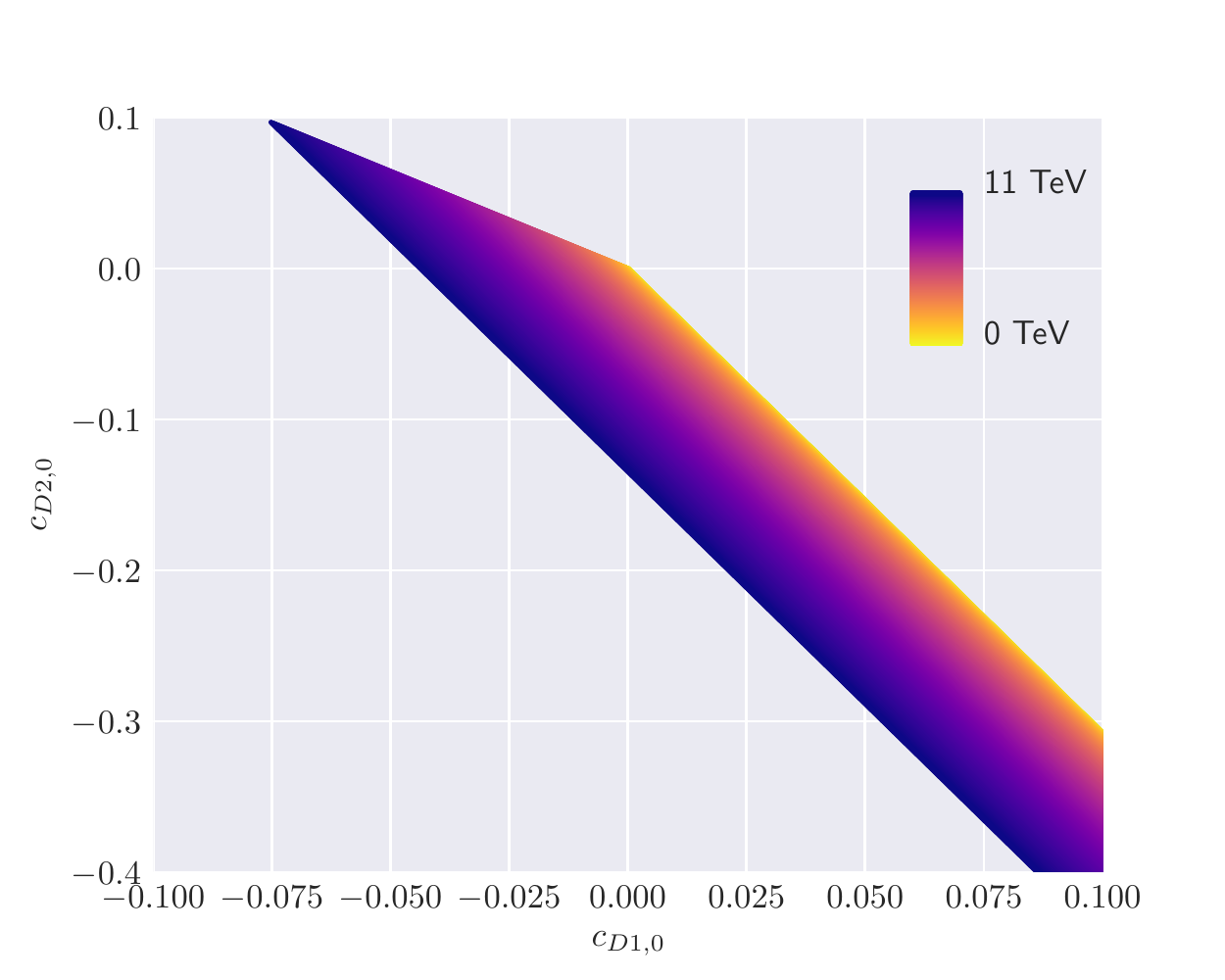}
 \caption{Skyrmion mass $M_{\text{Sk}}$ as a function of $c_{D1, 0}$ and $c_{D2, 0}$.}
 \label{fig:coeffs}
\end{figure}

It remains to study the skyrmions generated by $c_{D1,0}$ and $c_{D2,0}$. It turns out that both individually, as well as some of their linear combinations, generate meta-stable skyrmions. We parametrize the space of linear combinations with two parameters $e$ and $\theta$, with the former to be used as the corresponding parameter in the spherical ansatz:
\begin{equation}
 c_{D1,0} = \frac{\sqrt{2}}{16 e^2} \cos \theta,
 \qquad \qquad
 c_{D2,0} = \frac{\sqrt{2}}{16 e^2} \sin \theta.
\end{equation}
The Skyrme term is recovered for $\theta = 3 \pi / 4$. In terms of these parameters, the non-SM contribution to the radial energy reads
\begin{equation}
 c_{D1,0} \rho_{D1} + c_{D2, 0} \rho_{D2}
 =
 - \frac{\cos \theta}{4 r^2}
 \left[
  2 (f_1^2 + f_2^2) r^2 b^2
  + 2 (2 + \tan\theta) (f_1^2 + f_2^2)^2
  + (1 + \tan\theta) r^4 b^4
 \right]
\end{equation}
This is positive everywhere if and only if $\cos \theta \leq 0$ and $\tan\theta \geq -1$, or, equivalently $3 \pi / 4 \leq \theta \leq 3 \pi / 2$. Numerically, we find that skyrmions are stabilized in a slightly wider range:\footnote{The region determined by these values agrees with the one obtained in ref.~\cite{Ellis:2012cs} for the case in which both limit \textbf{A} and \textbf{B} are taken.}
\begin{figure}
 \centering
 \includegraphics[width=0.45\textwidth]{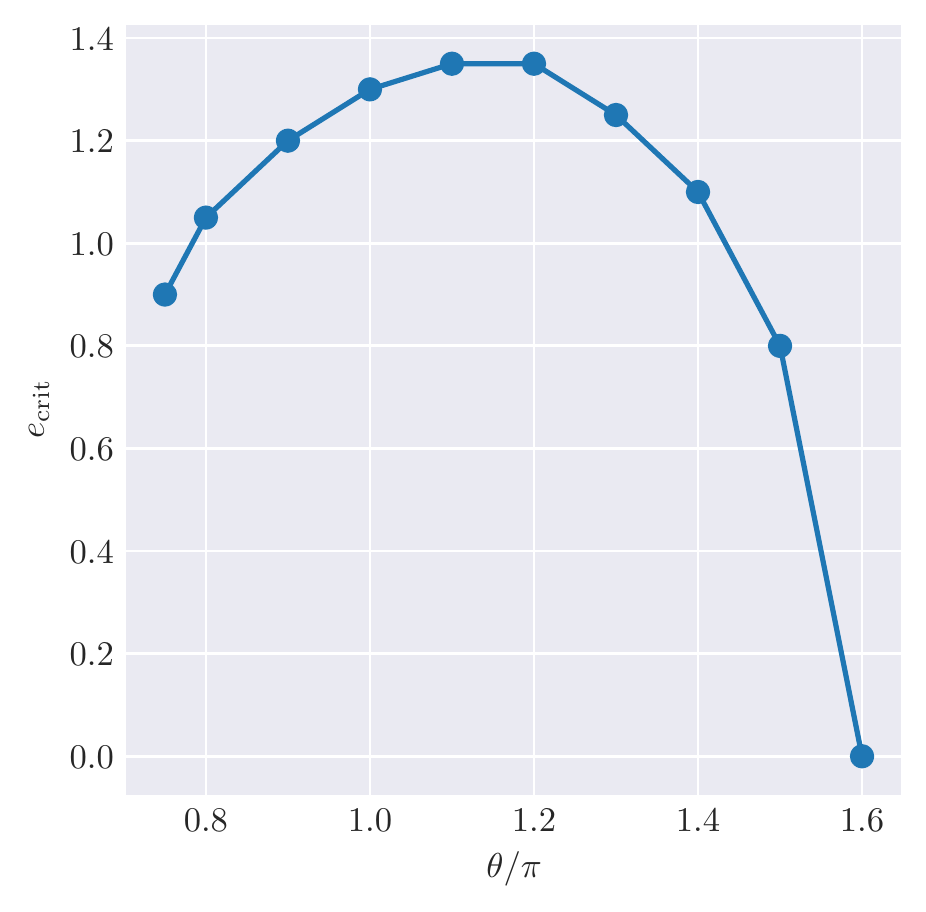}
 \includegraphics[width=0.45\textwidth]{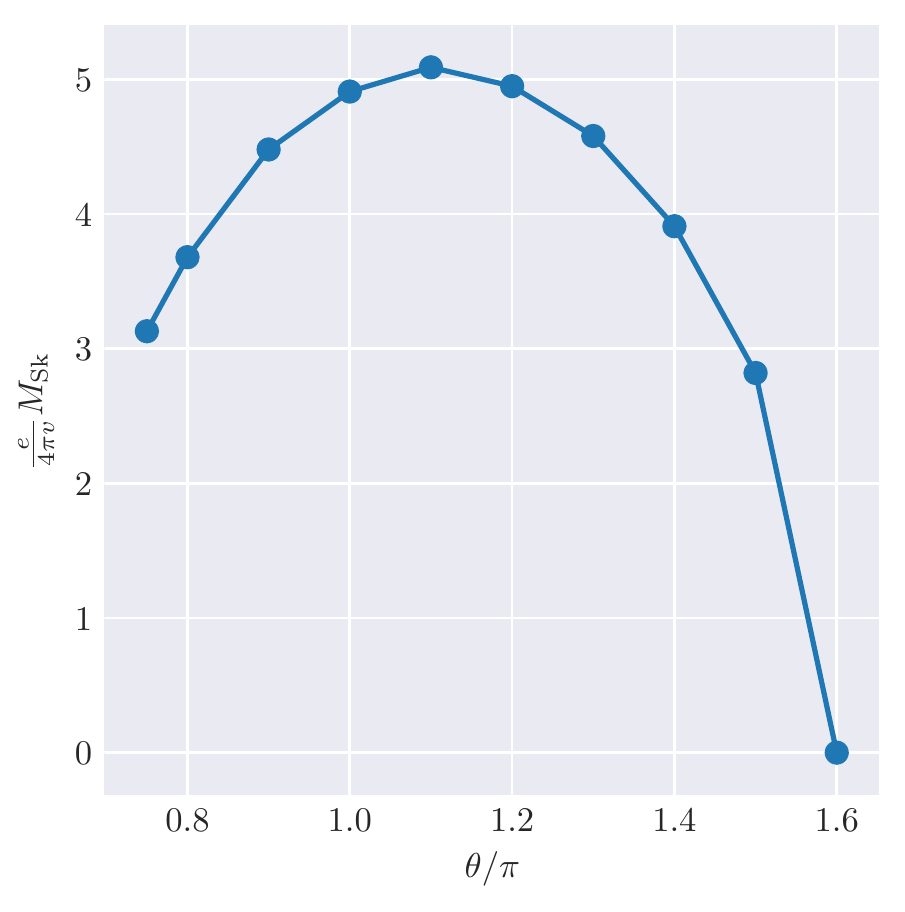}
 \caption{Left: $e_{\text{crit}}$ as a function of $\theta$.
  Right: skyrmion mass $M_{\text{Sk}}$ as a function of $\theta$, for $e = 3$.}
 \label{fig:eE-vs-theta}
\end{figure}
\begin{figure}
 \centering
 \includegraphics[width=0.55\textwidth]{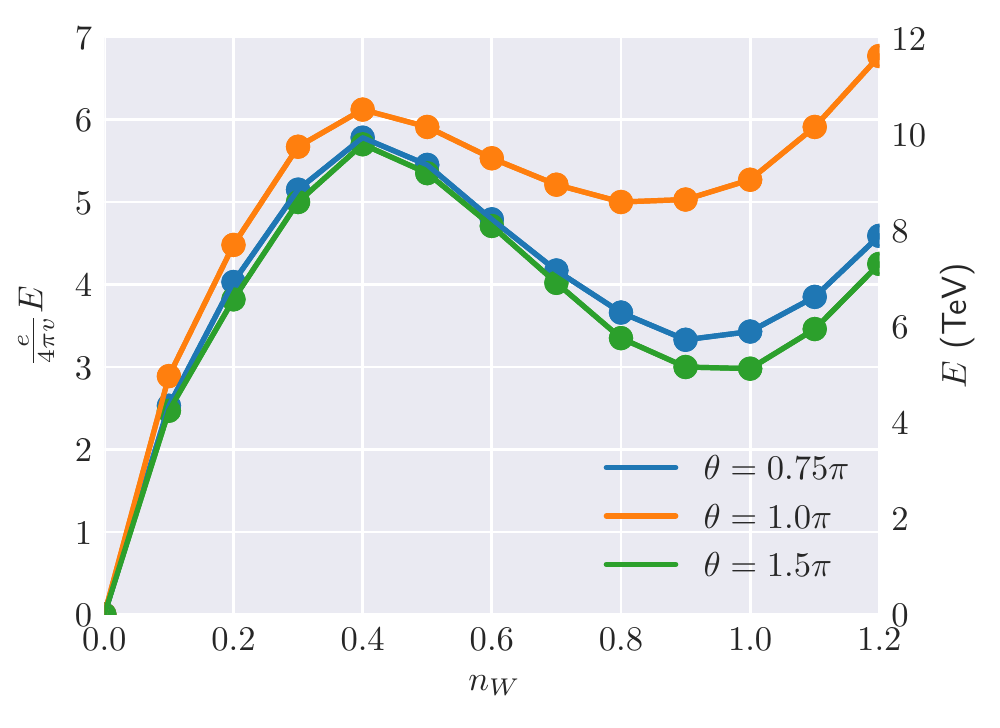}
 \caption{Minimal energy as a function of $n_W$, for $e = 1.8$ and $\theta = 3 \pi / 4, \pi, 3 \pi / 2$.}
 \label{fig:E-vs-nW-vs-theta}
\end{figure}
\begin{equation}
 0.71 \pi \simeq \theta_{\text{min}} \leq \theta \leq \theta_{\text{max}} \simeq 1.6 \pi,  
\end{equation}
for $e > e_{\text{crit}}(\theta)$, where $e_{\text{crit}}(\theta)$ is a $\theta$-dependent critical value of $e$, that we show on the left panel in figure~\ref{fig:eE-vs-theta}. The skyrmion mass also depends on $\theta$ for constant $e$, with $M_{\text{Sk}} = 0$ at $\theta = \theta_{\text{max}}$. We show this on the right panel of figure~\ref{fig:eE-vs-theta}. The normalized mass $e M_{\text{Sk}} / (4 \pi v)$ has little variation with $e$, as it happened for $\theta = 3 \pi / 4$. In figure~\ref{fig:E-vs-nW-vs-theta}, we display the energy profile for $e = 1.8 > \max_\theta e_{\text{crit}}(\theta)$, and different values of $\theta$. Finally, in figure~\ref{fig:coeffs} we show the region of $(c_{D1,0}, c_{D2,0})$ space where meta-stable skyrmions exist, and the values the masses of the skyrmions inside it, which are given approximately by
\begin{equation}
 M_{\text{Sk}} \simeq (\SI{30}{TeV}) \cdot
 \left[\tan(\theta_{\text{max}}) c_{D1,0} - c_{D2,0}\right]^{1/2}.
\end{equation}

The radius is similarly given by
\begin{equation}
 R_{\text{Sk}} \simeq (\SI{20}{TeV^{-1}}) \cdot
 \left[\tan(\theta_{\text{max}}) c_{D1,0} - c_{D2,0}\right]^{1/2}.
\end{equation}
The condition $e > e_{\text{crit}}(\theta)$ is just a $\theta$-independent upper bound on the skyrmion mass $M_{\text{Sk}} < \SI{11}{TeV}$. The region where skyrmions exist in the $(c_{D1,0}, c_{D2,0})$ plane is thus determined by
\begin{equation}
 c_{D2,0} < \tan(\theta_{\text{min}}) c_{D1,0},
 \qquad \qquad
 0 < \tan(\theta_{\text{max}}) c_{D1,0} - c_{D2,0} \lesssim 0.13.
\end{equation}

\begin{figure}
 \centering
 \includegraphics[width=0.49\textwidth]{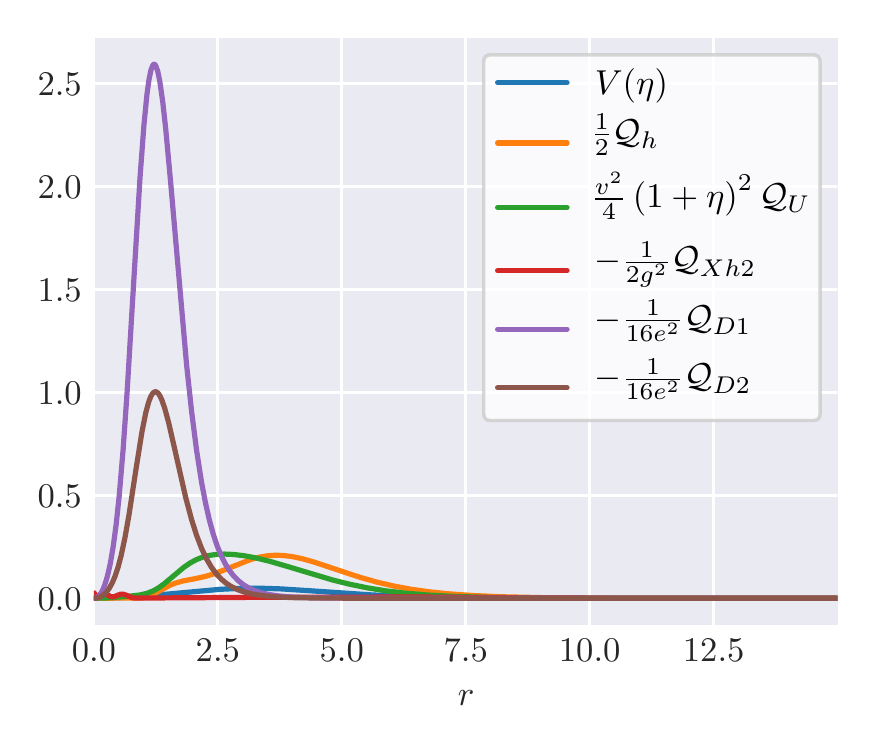}
 \includegraphics[width=0.49\textwidth]{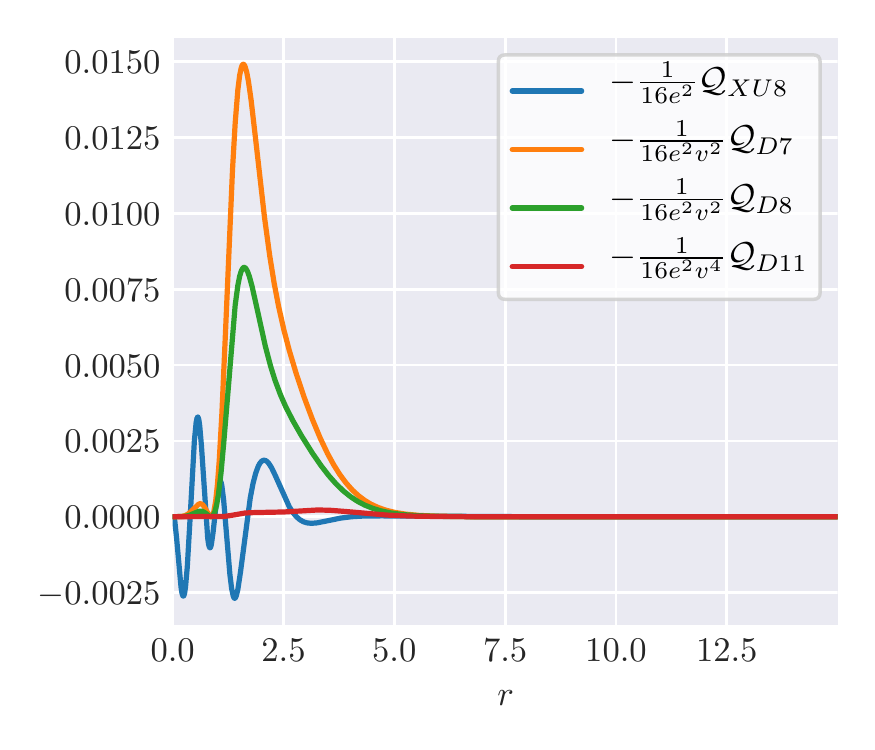}
 \caption{Radial energy densities $\rho_i$ in the skyrmion configuration with $\theta = 3 \pi / 4$ and $e = 1.8$. The operators in the left plot are included in the calculation of the skyrmion configuration. The ones in the right plot are computed once this configuration is obtained and fixed.}
 \label{fig:densities}
\end{figure}

Although the rest of the $c_{i, n}$ coefficients are not enough by themselves to stabilize skyrmions, they may have effects in the configurations generated by $c_{D1,0}$ and $c_{D2,0}$. Figure~\ref{fig:densities} shows the contribution of each $\mathcal{Q}_i$ to the energy density in the configuration with $\theta = 3 \pi / 4$ and $e = 1.8$. The contributions from the operators not included in the generation of the configuration are negligible compared to the energy. This means that whenever the $c_{i,n}$ coefficients are chosen so that their contribution is positive, they will not change the skyrmion configuration in a significant way. However, they might be chosen so that their contribution to the energy is arbitrarily negative, destabilizing the skyrmion. We find numerically that this happens when $c_{D8,0} = 1$, for example.

\section{Phenomenology}
\label{sec:pheno}

\subsection{Collider signals}

The process of electroweak skyrmion production is similar to the electroweak instanton, as it is a $B + L$ violating transition over a barrier of a few TeV. As such, it is expected to be exponentially suppressed, even at energies above the potential barrier~\cite{DasBakshi:2020ejz, Banks:1990zb}. Thus, it is unlikely that this process will take place at colliders. However, one can indirectly study the existence of skyrmions through other effects of the operators that generate them.

\begin{figure}
 \centering
 \includegraphics[width=0.8\textwidth]{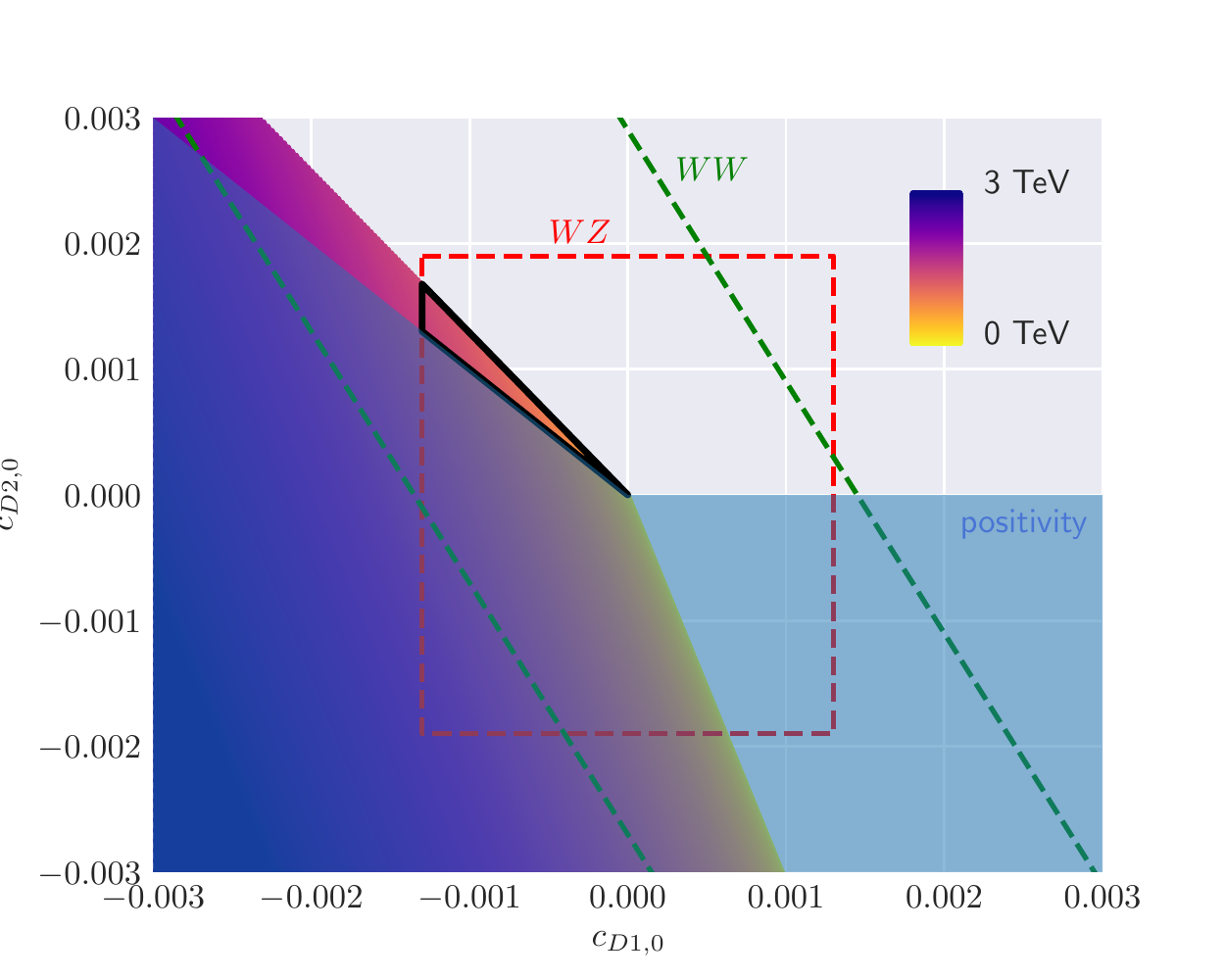}
 \caption{Dashed lines: 95\% CL limits on $c_{D1,0}$ and $c_{D2,0}$ at $\sqrt{s} = \SI{13}{TeV}$, $\int dt L = \SI{35.9}{fb^{-1}}$ from CMS~\cite{CMS:2019qfk}, using data from $WZ$ production (red) and $WW$ production (green). Transparent shaded blue region: excluded by positivity bounds. Color-gradient region: allowed values of the coefficients for the existence of skyrmions, from the numerical calculations in this work. The coloring represents the skyrmion mass. Solid black-line perimeter encloses the triangle of allowed values of the coefficients that support a skyrmion.}
 \label{fig:coeffs-limits}
\end{figure}

The two skyrmion-stabilizing operators $\mathcal{Q}_{D1}$ and $\mathcal{Q}_{D2}$ induce an anomalous quartic gauge coupling (aQGC) while preserving the SM triple gauge coupling. Most LHC searches for aQGC~\cite{CMS:2014mra,ATLAS:2015ify,ATLAS:2017vqm,ATLAS:2017bon,CMS:2019qfk,CMS:2020gfh,CMS:2020fqz} use a parametrization in terms of dimension-8 SMEFT operators which was first proposed in ref.~\cite{Eboli:2006wa}. This set of operators was corrected in ref.~\cite{Eboli:2016kko} by introducing missing operators and removing redundant ones in order for them to form a basis. The space of operators with four covariant derivatives was shown to have dimension 3. However, the experimental searches with the strongest constraint on this space~\cite{CMS:2019qfk,CMS:2020gfh} give their results in terms of only two operators, coming from an incomplete set of ref.~\cite{Eboli:2006wa}:
\begin{equation}
 \mathcal{L}_{S} =
 \frac{f_{S0}}{\Lambda^4} (D_\mu \phi^\dagger D_\nu \phi) (D^\mu \phi^\dagger D^\nu \phi)
 + \frac{f_{S1}}{\Lambda^4} (D_\mu \phi^\dagger D^\mu \phi) (D_\nu \phi^\dagger D^\nu \phi)
\end{equation}
Therefore, their results cannot be used in general to constrain the full 3-dimensional space of Wilson coefficients. Only when the measured final state uniquely selects one aQCG vertex ($WWWW$, $WWZZ$ or $ZZZZ$) can the results in the incomplete set be translated into the complete EFT basis, as shown in ref.~\cite{Rauch:2016pai}. Following this reference, we obtain limits over $c_{D1,0}$ and $c_{D2,0}$ (denoted $\alpha_5$ and $\alpha_4$ there) from the 95\% CL limits over $f_{S0}$ and $f_{S1}$ found in ref.~\cite{CMS:2020gfh} individually for $WW$ and $WZ$ production at $\sqrt{s} = \SI{13}{TeV}$ and $\int L dt = \SI{137}{fb^{-1}}$. $WW$ production comes from the $WWWW$ vertex, and the limits and conversion are given by
\begin{gather}
 \num{-2.7e-3}
 \leq
 2 c_{D1, 0} + c_{D2, 0} = \frac{v^4 f_{S1}}{8 \Lambda^4}
 \leq
 \num{2.9e-3},
 \\
 \num{-8.2e-3}
 \leq
 c_{D2, 0} = \frac{v^4 f_{S0}}{8 \Lambda^4}
 \leq
 \num{8.9e-3},
\end{gather}
whereas $WZ$ production comes from the $WWZZ$ vertex, they are
\begin{gather}
 \num{-1.3e-3}
 \leq
 c_{D1, 0} = \frac{v^4 f_{S1}}{16 \Lambda^4}
 \leq
 \num{1.3e-3},
 \\
 \num{-1.9e-3}
 \leq
 c_{D2, 0} = \frac{v^4 f_{S0}}{16 \Lambda^4}
 \leq
 \num{1.9e-3}.
\end{gather}
We show these limits in figure~\ref{fig:coeffs-limits}. We point out that the experimental bounds in ref.~\cite{ATLAS:2016nmw} are presented in terms of a basis for 2-dimensional custodial-invariant subspace of the 3-dimensional space of qQGC operators containing only covariant derivatives, and thus directly translatable into our setting. However, they are weaker than the ones we have obtained, and they are not shown in figure~\ref{fig:coeffs-limits}.

\subsection{Positivity bounds}

The space of Wilson coefficients can also be constrained theoretically by imposing general principles such as unitarity, locality and causality. The bounds obtained in this way are known as positivity bounds~\cite{Adams:2006sv}, and can be interpreted as necessary conditions for the existence of a UV completion to the EFT in question. In the HEFT, causality implies that~\cite{Distler:2006if, Fabbrichesi:2015hsa, Zhang:2018shp, Bi:2019phv}
\begin{equation}
 c_{D1, 0} + c_{D2, 0} > 0, \qquad \qquad c_{D2, 0} > 0.
\end{equation}
These inequalities also arise in the chiral Lagrangian without gauge bosons~\cite{Jenkins:2006ia}. The region excluded by them is shown in blue in figure~\ref{fig:coeffs-limits}. It follows that skyrmions can only exist in the angular region $\theta_{\text{min}} \leq \theta < 3 \pi / 4$. Combining this fact with the experimental limits gives an upper bound on the mass of the skyrmion:
\begin{equation}
 M_{\text{Sk}} \lesssim \SI{1.6}{TeV}.
\end{equation}

\subsection{Dark matter}


Similarly to skyrmion production, skyrmion decay is a $B + L$-violating process which expected to be exponentially suppressed. The skyrmion lifetime is thus likely longer than the age of the universe, opening the possibility of skyrmions being Dark Matter (DM) candidates. We use the following order-of-magnitude estimate of the freeze-out skyrmion density~\cite{Criado:2020zwu}
\begin{equation}
 \Omega_{\text{Sk}} h^2 \simeq \frac{\SI{3e-27}{cm^3 s^{-1}}}{\left<\sigma_{\text{ann}} \text{v}\right>},
 \qquad \qquad
 \sigma_{\text{ann}} \simeq \pi R_{\text{Sk}}^2,
 \qquad \qquad
 \text{v} \simeq 1/2.
\end{equation}
Requiring that the skyrmion density is at most the total DM density $\Omega_{\text{Sk}} h^2 \lesssim 0.1$ results in a lower limit for the skyrmion mass
\begin{equation}
 M_{\text{Sk}} \gtrsim \SI{60}{GeV}
\end{equation}
This limit would be saturated if skyrmions formed all of the DM.

\section{Conclusions}
\label{sec:conclusions}

We have studied the skyrmion configurations that arise in the HEFT. We have found that a meta-stable configuration with skyrmion number close to one exists whenever the coefficients $c_{D1, 0}$ and $c_{D2, 0}$ lie on the strip
\begin{equation*}
 c_{D2, 0} \leq \tan(\theta_{\text{min}}) c_{D1, 0},
 \qquad \qquad
 0 < \tan(\theta_{\text{max}}) c_{D1, 0} - c_{D2, 0} \lesssim 0.13.
\end{equation*}
The mass of this skyrmion is given by $M_{\text{Sk}} = (\SI{30}{TeV}) \cdot [\tan(\theta_{\text{max}}) c_{D1, 0} - c_{D2, 0}]^{1/2}$. It is separated from the trivial vacuum by an energy barrier of about $\SI{11}{TeV}$. This value also represents the maximal theoretical $M_{\text{Sk}}$, as above it the barrier disappears.

Since skyrmions are unlikely to be created at colliders, we have focused on the experimental signals of the operators that stabilize them. LHC searches for aQCG put bounds of order $\num{e-3}$ on both coefficients. Combining these bounds with positivity constraints, we have found that the allowed parameter space for skyrmions in the triangle
\begin{equation}
 1 < c_{D2, 0} / c_{D1, 0} \leq \tan(\theta_{\text{min}}),
 \qquad \qquad
 -\num{1.3e-3} \leq c_{D1, 0} \leq 0.
\end{equation}
This allowed us to obtain a stronger upper bound on the mass of the skyrmion, of about $\SI{1.6}{TeV}$.

Skyrmions are also expected to be long-lived, so they contribute to the DM density. By assuming that their abundance is generated by the freeze-out mechanism and adopting a simple approximation for the skyrmion annihilation cross-section, we have computed an order-of-magnitude lower bound on the skyrmion mass, of $\SI{60}{GeV}$.

\medskip

\appendix

\section{Numerical method}
\label{sec:numerical-method}

We use a neural network to model the function taking $r$ to $R(r) = \left(f_1(r), f_2(r), b(r), \eta(r)\right)$. We choose an architecture with sigmoid $\sigma(x) = (1 + e^{-x})^{-1}$ activation functions and two hidden layers, each having 5 units. That is, we parameterize
\begin{equation}
 R(r) = A_3 \circ \sigma \circ A_2 \circ \sigma \circ A_1(r)
\end{equation}
where $A_1 : \mathbb{R}^1 \to \mathbb{R}^{5}$, $A_2: \mathbb{R}^5 \to \mathbb{R}^5$, and $A_3 : \mathbb{R}^5 \to \mathbb{R}^4$ are affine transformations. The problem is to adjust the parameters of $A_1$, $A_2$ and $A_3$ to minimize the energy $E[R]$ while satisfying the boundary conditions $\operatorname{BC}[R] = 0$ and fixing $n_W[R]$ to some value $n_{W,0}$. We reformulate this problem as minimizing the functional
\begin{equation}
 L[R] = E[R] + \omega_{\text{BC}} \operatorname{BC}[R]^2 + \omega_{n_W} (n_W[R] - n_{W,0})^2,
\end{equation}
with the weights $\omega_i$ being sufficiently high. In practice, we set $\omega_{\text{BC}} = 10^3$ and $\omega_{n_W} = 10^4$. To perform the numerical minimization of $L$, the training of the network, we use the functional-minimization features of \texttt{Elvet}~\cite{Araz:2021hpx, Piscopo:2019txs}.

This procedure allows us to find the minimal energy $E$ for a fixed $n_W$. In order to find the local minimum near $n_W = 1$, which is the skyrmion, we first train the network with $n_W = 1$. Once the minimum under this condition is reached, we resume the training with this condition removed by setting $\omega_{n_W} = 0$. Since the network is already close to the local minimum, the training cannot overcome the finite barrier, so it can only take the network into the skyrmion configuration.

We check that the configurations found in this way are compatible with Derrick's argument. We first split the energy as
\begin{equation}
 E = \sum_i E_i,
 \qquad \qquad
 E_i = \frac{4 \pi v}{e} \int_0^\infty dr F_i(\eta) \rho_i.
\end{equation}
In terms of the functions defined here, a spatial scale transformation is given as $r \to \lambda r$, $b \to b / \lambda$. If one applies such transformation to a local minimum, the energy must satisfy:
\begin{align}
 0 = \left. \frac{d E}{d \lambda} \right|_{\lambda = 1}
 &=
  3 E_1 + (E_h + E_U) - \sum_i E_i,
  \label{eq:derrick}
\end{align}
where $i$ runs over rest of operators in table~\ref{tab:operators}. Numerically, we get
\begin{equation}
 \frac{1}{E} \left. \frac{d E}{d \lambda} \right|_{\lambda = 1} < 1\%.
\end{equation}

In order to find the critical value $e_{\text{crit}}$, we note that for $e > e_{\text{crit}}$ the height of the local minimum must be lower than the height of the barrier, so at $e = e_{\text{crit}}$ they must become of the same height. For small $e$, the barrier is found at around $n_W = 0.4$ and the local minimum is close to $n_W = 0.8$. We can thus obtain $e_{\text{crit}}$ approximately as the value of $e$ that minimizes $|E(n_W = 0.4) - E(n_W = 0.8)|$. We search for this value in steps of 0.05.

\bibliographystyle{JHEP}
\bibliography{references}

\providecommand{\href}[2]{#2}\begingroup\raggedright\begin{thebibliography}{10}

\bibitem{Skyrme:1961vq}
T.H.R.~Skyrme, \emph{{A Nonlinear field theory}},
  \href{https://doi.org/10.1098/rspa.1961.0018}{\emph{Proc. Roy. Soc. Lond. A}
  {\bfseries 260} (1961) 127}.

\bibitem{Witten:1979kh}
E.~Witten, \emph{{Baryons in the 1/n Expansion}},
  \href{https://doi.org/10.1016/0550-3213(79)90232-3}{\emph{Nucl. Phys. B}
  {\bfseries 160} (1979) 57}.

\bibitem{Adkins:1983ya}
G.S.~Adkins, C.R.~Nappi and E.~Witten, \emph{{Static Properties of Nucleons in
  the Skyrme Model}},
  \href{https://doi.org/10.1016/0550-3213(83)90559-X}{\emph{Nucl. Phys. B}
  {\bfseries 228} (1983) 552}.

\bibitem{Ellis:2012cs}
J.~Ellis, M.~Karliner and M.~Praszalowicz, \emph{{Generalized Skyrmions in QCD
  and the Electroweak Sector}},
  \href{https://doi.org/10.1007/JHEP03(2013)163}{\emph{JHEP} {\bfseries 03}
  (2013) 163} [\href{https://arxiv.org/abs/1209.6430}{{\ttfamily 1209.6430}}].

\bibitem{Ambjorn:1984bb}
J.~Ambjorn and V.A.~Rubakov, \emph{{Classical Versus Semiclassical Electroweak
  Decay of a Techniskyrmion}},
  \href{https://doi.org/10.1016/0550-3213(85)90403-1}{\emph{Nucl. Phys. B}
  {\bfseries 256} (1985) 434}.

\bibitem{Criado:2020zwu}
J.C.~Criado, V.V.~Khoze and M.~Spannowsky, \emph{{The Emergence of Electroweak
  Skyrmions through Higgs Bosons}},
  \href{https://doi.org/10.1007/JHEP03(2021)162}{\emph{JHEP} {\bfseries 03}
  (2021) 162} [\href{https://arxiv.org/abs/2012.07694}{{\ttfamily
  2012.07694}}].

\bibitem{Derrick:1964ww}
G.H.~Derrick, \emph{{Comments on nonlinear wave equations as models for
  elementary particles}}, \href{https://doi.org/10.1063/1.1704233}{\emph{J.
  Math. Phys.} {\bfseries 5} (1964) 1252}.

\bibitem{ATLAS:2012yve}
{\scshape ATLAS} collaboration, \emph{{Observation of a new particle in the
  search for the Standard Model Higgs boson with the ATLAS detector at the
  LHC}}, \href{https://doi.org/10.1016/j.physletb.2012.08.020}{\emph{Phys.
  Lett. B} {\bfseries 716} (2012) 1}
  [\href{https://arxiv.org/abs/1207.7214}{{\ttfamily 1207.7214}}].

\bibitem{CMS:2012qbp}
{\scshape CMS} collaboration, \emph{{Observation of a New Boson at a Mass of
  125 GeV with the CMS Experiment at the LHC}},
  \href{https://doi.org/10.1016/j.physletb.2012.08.021}{\emph{Phys. Lett. B}
  {\bfseries 716} (2012) 30} [\href{https://arxiv.org/abs/1207.7235}{{\ttfamily
  1207.7235}}].

\bibitem{Buchalla:2018yce}
G.~Buchalla, M.~Capozi, A.~Celis, G.~Heinrich and L.~Scyboz, \emph{{Higgs boson
  pair production in non-linear Effective Field Theory with full
  $m_t$-dependence at NLO QCD}},
  \href{https://doi.org/10.1007/JHEP09(2018)057}{\emph{JHEP} {\bfseries 09}
  (2018) 057} [\href{https://arxiv.org/abs/1806.05162}{{\ttfamily
  1806.05162}}].

\bibitem{Buchalla:2013rka}
G.~Buchalla, O.~Cat\`a and C.~Krause, \emph{{Complete Electroweak Chiral
  Lagrangian with a Light Higgs at NLO}},
  \href{https://doi.org/10.1016/j.nuclphysb.2014.01.018}{\emph{Nucl. Phys. B}
  {\bfseries 880} (2014) 552}
  [\href{https://arxiv.org/abs/1307.5017}{{\ttfamily 1307.5017}}].

\bibitem{Kitano:2016ooc}
R.~Kitano and M.~Kurachi, \emph{{Electroweak-Skyrmion as Topological Dark
  Matter}}, \href{https://doi.org/10.1007/JHEP07(2016)037}{\emph{JHEP}
  {\bfseries 07} (2016) 037}
  [\href{https://arxiv.org/abs/1605.07355}{{\ttfamily 1605.07355}}].

\bibitem{Kitano:2017zqw}
R.~Kitano and M.~Kurachi, \emph{{More on Electroweak-Skyrmion}},
  \href{https://doi.org/10.1007/JHEP04(2017)150}{\emph{JHEP} {\bfseries 04}
  (2017) 150} [\href{https://arxiv.org/abs/1703.06397}{{\ttfamily
  1703.06397}}].

\bibitem{Hamada:2021oqm}
Y.~Hamada, R.~Kitano and M.~Kurachi, \emph{{Electroweak-Skyrmion as Asymmetric
  Dark Matter}},  \href{https://arxiv.org/abs/2108.12185}{{\ttfamily
  2108.12185}}.

\bibitem{DasBakshi:2020ejz}
S.~Das~Bakshi, J.~Chakrabortty, C.~Englert, M.~Spannowsky and P.~Stylianou,
  \emph{{$CP$ violation at ATLAS in effective field theory}},
  \href{https://doi.org/10.1103/PhysRevD.103.055008}{\emph{Phys. Rev. D}
  {\bfseries 103} (2021) 055008}
  [\href{https://arxiv.org/abs/2009.13394}{{\ttfamily 2009.13394}}].

\bibitem{Banks:1990zb}
T.~Banks, G.R.~Farrar, M.~Dine, D.~Karabali and B.~Sakita, \emph{{WEAK
  INTERACTIONS ARE WEAK AT HIGH-ENERGIES}},
  \href{https://doi.org/10.1016/0550-3213(90)90376-O}{\emph{Nucl. Phys. B}
  {\bfseries 347} (1990) 581}.

\bibitem{CMS:2014mra}
{\scshape CMS} collaboration, \emph{{Study of vector boson scattering and
  search for new physics in events with two same-sign leptons and two jets}},
  \href{https://doi.org/10.1103/PhysRevLett.114.051801}{\emph{Phys. Rev. Lett.}
  {\bfseries 114} (2015) 051801}
  [\href{https://arxiv.org/abs/1410.6315}{{\ttfamily 1410.6315}}].

\bibitem{ATLAS:2015ify}
{\scshape ATLAS} collaboration, \emph{{Evidence of
  W\ensuremath{\gamma}\ensuremath{\gamma} Production in pp Collisions at s=8
  TeV and Limits on Anomalous Quartic Gauge Couplings with the ATLAS
  Detector}}, \href{https://doi.org/10.1103/PhysRevLett.115.031802}{\emph{Phys.
  Rev. Lett.} {\bfseries 115} (2015) 031802}
  [\href{https://arxiv.org/abs/1503.03243}{{\ttfamily 1503.03243}}].

\bibitem{ATLAS:2017vqm}
{\scshape ATLAS} collaboration, \emph{{Studies of $Z\gamma$ production in
  association with a high-mass dijet system in $pp$ collisions at $\sqrt{s}=$ 8
  TeV with the ATLAS detector}},
  \href{https://doi.org/10.1007/JHEP07(2017)107}{\emph{JHEP} {\bfseries 07}
  (2017) 107} [\href{https://arxiv.org/abs/1705.01966}{{\ttfamily
  1705.01966}}].

\bibitem{ATLAS:2017bon}
{\scshape ATLAS} collaboration, \emph{{Study of $WW\gamma$ and $WZ\gamma$
  production in $pp$ collisions at $\sqrt{s} = 8$ TeV and search for anomalous
  quartic gauge couplings with the ATLAS experiment}},
  \href{https://doi.org/10.1140/epjc/s10052-017-5180-3}{\emph{Eur. Phys. J. C}
  {\bfseries 77} (2017) 646}
  [\href{https://arxiv.org/abs/1707.05597}{{\ttfamily 1707.05597}}].

\bibitem{CMS:2019qfk}
{\scshape CMS} collaboration, \emph{{Search for anomalous electroweak
  production of vector boson pairs in association with two jets in
  proton-proton collisions at 13 TeV}},
  \href{https://doi.org/10.1016/j.physletb.2019.134985}{\emph{Phys. Lett. B}
  {\bfseries 798} (2019) 134985}
  [\href{https://arxiv.org/abs/1905.07445}{{\ttfamily 1905.07445}}].

\bibitem{CMS:2020gfh}
{\scshape CMS} collaboration, \emph{{Measurements of production cross sections
  of WZ and same-sign WW boson pairs in association with two jets in
  proton-proton collisions at $\sqrt{s} =$ 13 TeV}},
  \href{https://doi.org/10.1016/j.physletb.2020.135710}{\emph{Phys. Lett. B}
  {\bfseries 809} (2020) 135710}
  [\href{https://arxiv.org/abs/2005.01173}{{\ttfamily 2005.01173}}].

\bibitem{CMS:2020fqz}
{\scshape CMS} collaboration, \emph{{Evidence for electroweak production of
  four charged leptons and two jets in proton-proton collisions at $\sqrt {s}$
  = 13 TeV}}, \href{https://doi.org/10.1016/j.physletb.2020.135992}{\emph{Phys.
  Lett. B} {\bfseries 812} (2021) 135992}
  [\href{https://arxiv.org/abs/2008.07013}{{\ttfamily 2008.07013}}].

\bibitem{Eboli:2006wa}
O.J.P.~Eboli, M.C.~Gonzalez-Garcia and J.K.~Mizukoshi, \emph{{p p
  ---\ensuremath{>} j j e+- mu+- nu nu and j j e+- mu-+ nu nu at O(
  alpha(em)**6) and O(alpha(em)**4 alpha(s)**2) for the study of the quartic
  electroweak gauge boson vertex at CERN LHC}},
  \href{https://doi.org/10.1103/PhysRevD.74.073005}{\emph{Phys. Rev. D}
  {\bfseries 74} (2006) 073005}
  [\href{https://arxiv.org/abs/hep-ph/0606118}{{\ttfamily hep-ph/0606118}}].

\bibitem{Eboli:2016kko}
O.J.P.~\'Eboli and M.C.~Gonzalez-Garcia, \emph{{Classifying the bosonic quartic
  couplings}}, \href{https://doi.org/10.1103/PhysRevD.93.093013}{\emph{Phys.
  Rev. D} {\bfseries 93} (2016) 093013}
  [\href{https://arxiv.org/abs/1604.03555}{{\ttfamily 1604.03555}}].

\bibitem{Rauch:2016pai}
M.~Rauch, \emph{{Vector-Boson Fusion and Vector-Boson Scattering}},
  \href{https://arxiv.org/abs/1610.08420}{{\ttfamily 1610.08420}}.

\bibitem{ATLAS:2016nmw}
{\scshape ATLAS} collaboration, \emph{{Search for anomalous electroweak
  production of $WW/WZ$ in association with a high-mass dijet system in $pp$
  collisions at $\sqrt{s}=8$ TeV with the ATLAS detector}},
  \href{https://doi.org/10.1103/PhysRevD.95.032001}{\emph{Phys. Rev. D}
  {\bfseries 95} (2017) 032001}
  [\href{https://arxiv.org/abs/1609.05122}{{\ttfamily 1609.05122}}].

\bibitem{Adams:2006sv}
A.~Adams, N.~Arkani-Hamed, S.~Dubovsky, A.~Nicolis and R.~Rattazzi,
  \emph{{Causality, analyticity and an IR obstruction to UV completion}},
  \href{https://doi.org/10.1088/1126-6708/2006/10/014}{\emph{JHEP} {\bfseries
  10} (2006) 014} [\href{https://arxiv.org/abs/hep-th/0602178}{{\ttfamily
  hep-th/0602178}}].

\bibitem{Distler:2006if}
J.~Distler, B.~Grinstein, R.A.~Porto and I.Z.~Rothstein, \emph{{Falsifying
  Models of New Physics via WW Scattering}},
  \href{https://doi.org/10.1103/PhysRevLett.98.041601}{\emph{Phys. Rev. Lett.}
  {\bfseries 98} (2007) 041601}
  [\href{https://arxiv.org/abs/hep-ph/0604255}{{\ttfamily hep-ph/0604255}}].

\bibitem{Fabbrichesi:2015hsa}
M.~Fabbrichesi, M.~Pinamonti, A.~Tonero and A.~Urbano, \emph{{Vector boson
  scattering at the LHC: A study of the WW $\to$ WW channels with the Warsaw
  cut}}, \href{https://doi.org/10.1103/PhysRevD.93.015004}{\emph{Phys. Rev. D}
  {\bfseries 93} (2016) 015004}
  [\href{https://arxiv.org/abs/1509.06378}{{\ttfamily 1509.06378}}].

\bibitem{Zhang:2018shp}
C.~Zhang and S.-Y.~Zhou, \emph{{Positivity bounds on vector boson scattering at
  the LHC}}, \href{https://doi.org/10.1103/PhysRevD.100.095003}{\emph{Phys.
  Rev. D} {\bfseries 100} (2019) 095003}
  [\href{https://arxiv.org/abs/1808.00010}{{\ttfamily 1808.00010}}].

\bibitem{Bi:2019phv}
Q.~Bi, C.~Zhang and S.-Y.~Zhou, \emph{{Positivity constraints on aQGC: carving
  out the physical parameter space}},
  \href{https://doi.org/10.1007/JHEP06(2019)137}{\emph{JHEP} {\bfseries 06}
  (2019) 137} [\href{https://arxiv.org/abs/1902.08977}{{\ttfamily
  1902.08977}}].

\bibitem{Jenkins:2006ia}
A.~Jenkins and D.~O'Connell, \emph{{The Story of O: Positivity constraints in
  effective field theories}},
  \href{https://arxiv.org/abs/hep-th/0609159}{{\ttfamily hep-th/0609159}}.

\bibitem{Araz:2021hpx}
J.Y.~Araz, J.C.~Criado and M.~Spannowsky, \emph{{Elvet -- a neural
  network-based differential equation and variational problem solver}},
  \href{https://arxiv.org/abs/2103.14575}{{\ttfamily 2103.14575}}.

\bibitem{Piscopo:2019txs}
M.L.~Piscopo, M.~Spannowsky and P.~Waite, \emph{{Solving differential equations
  with neural networks: Applications to the calculation of cosmological phase
  transitions}}, \href{https://doi.org/10.1103/PhysRevD.100.016002}{\emph{Phys.
  Rev. D} {\bfseries 100} (2019) 016002}
  [\href{https://arxiv.org/abs/1902.05563}{{\ttfamily 1902.05563}}].

\end{thebibliography}\endgroup

\end{document}